\documentclass[11pt,a4paper]{revtex4-1}
\usepackage{amsmath,amssymb,enumerate,epsfig,pifont}

\newtheorem{definition}{Definition}
\newtheorem{proof}{Proof}

\def\A{\mathcal{A}}

\def\R{\mathbb{R}}
\def\C{\mathbb{C}}
\def\N{\mathbb{N}}
\def\Z{\mathbb{Z}}
\def\K{\mathbb{K}}
\def\g{\mathfrak{g}}

\def\p{\partial}

\def\al{\alpha}
\def\CC{\mathbf{C}}

\def\KK{\mathbf{K}}

\def\wtbfC{\widetilde{\mathbf{C}}}

\begin{document}

\title{Generalization of Weyl realization to a class of  Lie superalgebras}

\author{Stjepan Meljanac}
\email{meljanac@irb.hr}
\affiliation{Rudjer Bo\v{s}kovi\'{c} Institute, Theoretical Physics Division, Bijeni\v{c}ka c. 54, HR 10002 Zagreb, Croatia}

\author{Sa\v{s}a Kre\v{s}i\'{c}--Juri\'{c}}
\email{skresic@pmfst.hr}
\affiliation{Faculty of Science, University of Split, Rudjera Bo\v{s}kovi\'{c}a 33, 21000 Split, Croatia}

\author{Danijel Pikuti\'{c}}
\email{dpikutic@irb.hr}
\affiliation{Rudjer Bo\v{s}kovi\'{c} Institute, Theoretical Physics Division, Bijeni\v{c}ka c. 54, HR 10002 Zagreb, Croatia}

\begin{abstract}
This paper generalizes Weyl realization to a class of Lie superalgebras $\g=\g_0\oplus \g_1$ satisfying
$[\g_1,\g_1]=\{0\}$. First, we give a novel proof of the Weyl realization of a Lie algebra $\g_0$ by deriving
a functional equation for the function that defines the realization. We show that this equation has a unique
solution given by the generating function for the Bernoulli numbers. This method is then generalized
to Lie superalgebras of the above type.
\end{abstract}


\maketitle


\section{Introduction}
\label{introduction}

Realizations of Lie algebras have been used extensively in the study of noncommutative (NC) spaces and their deformed symmetries. The study of NC spaces is
motivated by physical evidence that the classical notion of a point at the Planck scale $l_P$ ($l_P=\sqrt{G\hbar/c^3} \approx 1.62\times 10^{-35}\, m$)
is no longer valid due to quantum fluctuations. Einstein's theory of gravity coupled with Heisenberg's uncertainty principle suggests that
space--time coordinates should satisfy uncertainly relations $\Delta x_\mu \Delta x_\nu \geq l_P^2\,$ \cite{DFR1, DFR2}. This requires a modification of our usual notion
of space--time as a continuum. In one of the possible approaches to a description of space--time at the Planck scale, the space--time coordinates are
replaced by a noncommutative algebra of operators $\hat x_1, \hat x_2, \ldots, \hat x_n$. Noncommutativity is introduced via commutation relations $[\hat x_\mu,
\hat x_\nu]=\theta_{\mu\nu}(\hat x)$ where $\theta_{\mu\nu}(\hat x)$ depends on a deformation parameter $h\in \R$ such that in the classical limit
$\theta_{\mu\nu}(\hat x)\to 0$ as $h\to 0$. The three most commonly studied types of NC spaces in the literature are the Moyal space \cite{Moyal, Groenewold}, $\kappa$--deformed
space \cite{LNR, Majid-Ruegg} and Snyder space \cite{Snyder}.  Noncommutative spaces have applications in quantum gravity \cite{Amelino-Camelia-3},
quantum field theory \cite{Daszkiewicz-1, Govindarajan, Amelino-Camelia-4} and deformed special relativity theories \cite{Kowalski-Glikman-1,
Kowalski-Glikman-2}. A class of Lie deformed Minkowski
spaces was considered in Refs. \cite{Govindarajan2, IJMPA2014, Mercati, LW, Kumar1703}.
A review of applications of NC spaces in physics can be found in Refs. \cite{Li} and \cite{Aschieri}.

Introduction of NC coordinates $\hat x_1, \hat x_2, \ldots, \hat x_n$ for description of space--time at the Planck scale requires the study of differential
geometry on quantum spaces as a proper mathematical tool for formulation of physical theories at such scale. The general theory of differential calculus
on quantum spaces was initiated by Woronowicz in Ref. \cite{Woronowicz}, and by now has been investigated by many authors from different points of view in Refs.
\cite{Vladimirov, Schupp, Landi, Sitarz, Gonera, Mercati2016, Meljanac-1, SKJ-FODC}.
A bicovariant differential calculus on Lie algebra type NC spaces was investigated in Ref. \cite{SKJ-FODC}. In this paper the authors use realizations of Lie
superalgebras by formal power series in the Weyl superalgebra (Clifford--Weyl algebra) in order to construct geometric objects as deformations of the
corresponding notions on the Euclidean space.

In the present work we prove a conjecture from Ref. \cite{SKJ-FODC} which states that the Weyl symmetric realization of a Lie algebra $\g_0$
can be generalized to Lie superalgebras $\g=\g_0\oplus \g_1$ satisfying $[\g_1,\g_1]=\{0\}$ where $[\cdot\, ,\, \cdot]$ is the supercommutator in $\g$.
For details on the Weyl realization of $\g_0$ see Refs. \cite{Durov} and \cite{SKJ}.
The evidence for this conjecture was found indirectly by using an automorphism of the Weyl superalgebra and comparing the power series expansion of
two different realizations of $\g$ to second order in the structure constants of $\g$. Here we give an elegant proof of this conjecture using the
functional equation satisfied by the generating function for the Bernoulli numbers. As a special case, this also gives a shorter proof of the Weyl
symmetric realization of $\g_0$ found in Refs. \cite{Durov} and \cite{SKJ}.

The plan of the paper is as follows. In Section 2 we give a novel proof of the Weyl symmetric realization of a Lie algebra $\g_0$ by formal power series
in the Weyl algebra. The proof is based on deriving a functional equation for the function that defines the realization of the generators of $\g_0$. It is
shown that this equation has a unique solution given by the generating function for the Bernoulli numbers $f(t)=t/(1-e^{-t})$. In Section 3 the proof is
generalized to Lie superalgebras $\g=\g_0\oplus \g_1$ such that $[\g_1, \g_1]=\{0\}$. Here, the generators of $\g$ are realized by formal power series in the
Weyl superalgebra and the condition $[\g_1, \g_1]=\{0\}$ ensures that the the function defining the realization satisfies the same functional equation as in Section 2.
Thus, the realization of $\g$ is also given in terms of the generating function $f(t)=t/(1-e^{-t})$. This establishes the conjecture stated in Ref.
\cite{SKJ-FODC}. The Appendix provides alternative methods for solving the functional equation for $f(t)$.

\section{The Weyl formula for Lie algebras}
\label{sec-2}

Let $\g_0$ be a Lie algebra over $\K$ ($\K=\R$ or $\C$) with ordered basis $X_1, X_2, \ldots, X_n$ satisfying the Lie bracket relations
\begin{equation}\label{2-1}
[X_\mu,X_\nu]=\sum_{\al=1}^n C_{\mu\nu\al} X_\al.
\end{equation}
The structure constants $C_{\mu\nu\al}$ satisfy $C_{\mu\nu\al}=-C_{\nu\mu\al}$ and the Jacobi identity
\begin{equation}\label{Jacobi}
\sum_{\rho=1}^n \big(C_{\mu\al\rho} C_{\rho\beta\nu}+C_{\al\beta\rho} C_{\rho\mu\nu} + C_{\beta\mu\rho} C_{\rho\al\nu}\big)=0.
\end{equation}
Our goal is to find a realization of the generators $X_\mu$ by formal power series in a semicompleted Weyl algebra $\A_n$. Recall that $\A_n$ is a unital
associative algebra generated by $x_\mu,\p_\mu$, $1\leq \mu\leq n$, satisfying the commutation relations $[x_\mu,x_\nu]=[\p_\mu,\p_\nu]=0$
and $[\p_\mu,x_\nu]=\delta_{\mu\nu}$. The algebra $\A_n$ has a faithful representation on the space of polynomials $\K[x_1, \ldots, x_n]$ where $x_\mu$
denotes the multiplication operator by $x_\mu$ and $\p_\mu$ is the partial derivative $\p_\mu=\frac{\p}{\p x_\mu}$. We define $\hat \A_n$ to be the
semicompletion of $\A_n$ containing formal power series in $\p_1, \ldots, \p_n$, but only polynomial expressions in $x_1, \ldots, x_n$. For the purpose of
our discussion we assume that the structure constants of $\g_0$ depend on a deformation parameter $h\in \R$ such that $\lim_{h\to 0} C_{\mu\nu\al}=0$.
This is always possible by simply rescaling the structure constants $C_{\mu\nu\al} \mapsto hC_{\mu\nu\al}$. The deformation parameter $h$ is introduced
because the enveloping algebra $U(\g_0)$ is interpreted as a space of functions on the underlying NC space with coordinates
$X_1, X_2, \ldots, X_n$ such that in the classical limit as $h\to 0$ we have $X_\mu X_\nu = X_\nu X_\mu$. Here, to simplify the notation we identify $X_\mu$
with its canonical image in $U(\g_0)$.

\begin{definition}
A realization of $\g_0$ is a Lie algebra monomorphism $\varphi \colon \g_0\to \hat \A_n$ defined by
\begin{equation}\label{2.3A}
\varphi(X_\mu) = \sum_{\al=1}^n x_\al \, \varphi_{\al\mu}(\p)
\end{equation}
where $\varphi_{\al\mu}(\p)$ is a formal power series in $\p_1, \ldots, \p_n$ depending on $h\in \R$ such that\\ $\lim_{h\to 0} \varphi_{\al\mu}(\p)=\delta_{\al\mu}$.
\end{definition}
Let us denote the realization $\varphi(X_\mu)$ by $\hat X_\mu$. In the classical limit we have $\lim_{h\to 0} \hat X_\mu = x_\mu$, hence $\hat X_\mu$ can be interpreted as a deformation
of the commutative coordinate $x_\mu$. The map $\varphi$ extends to a unique homomorphism of associative algebras $\varphi \colon U(\g_0)\to \hat \A_n$, hence $\hat X_\mu$ satisfy
the commutation relations $\hat X_\mu \hat X_\nu-\hat X_\nu \hat X_\mu = \sum_{\al=1}^n C_{\mu\al\nu} \hat X_\al$. In this section we consider a realization of the form
\begin{equation}\label{2.4A}
\hat X_\mu = \sum_{\al=1}^n x_\al f(\CC)_{\mu\al}
\end{equation}
where $\CC$ is the $n\times n$ matrix of differential operators $\CC_{\mu\nu}=\sum_{\al=1}^n C_{\mu\al\nu} \p_\al$ and $f(t)$ is a real analytic function such
that $f(0)=1$. The condition $f(0)=1$ ensures that $\lim_{h\to 0}\hat X_\mu = x_\mu$.
Starting with the commutation relations for $\hat X_\mu$, we shall derive a functional equation for $f(t)$. We will show that if $f(0)=1$, then $f(t)$ is
the generating function for the Bernoulli numbers $B_k$,
\begin{equation}\label{Bernoulli}
f(t)=\frac{t}{1-e^{-t}} = \sum_{k=0}^\infty B_k \frac{(-1)^k}{k!} t^k.
\end{equation}
First, let us prove some identities required to derive the functional equation for $f(t)$. From antisymmetry of the structure constants $C_{\mu\nu\al}=-C_{\nu\mu\al}$
and the Jacobi identity \eqref{Jacobi} we obtain
\begin{equation}\label{2-6}
\sum_{\al=1}^n C_{\mu\nu\al} \CC_{\al\lambda} = \sum_{\al,\beta=1}^n \big(\CC_{\mu\al}\, \delta_{\beta\nu}+\CC_{\nu\beta}\, \delta_{\al\mu}\big) C_{\al\beta\lambda}.
\end{equation}
It is convenient to introduce the following notation for two matrices $A$ and $B$,
\begin{equation}
(A\otimes B)_{\mu\al\nu\beta} = A_{\mu\al} B_{\nu\beta}.
\end{equation}
We also define $(A\otimes B)(C\otimes D)=AC\otimes BD$ whenever the matrix multiplication is defined.
In this notation Eq. \eqref{2-6} can be written as
\begin{equation}
\sum_{\al=1}^n C_{\mu\nu\al} \CC_{\al\lambda} = \sum_{\al,\beta=1}^n \big(\CC\otimes I + I \otimes \CC\big)_{\mu\al\nu\beta}\, C_{\al\beta\lambda},
\end{equation}
where $I$ is the $n\times n$ unit matrix. This readily generalizes to
\begin{equation}\label{2-9}
\sum_{\al=1}^n C_{\mu\nu\al} (\CC^m)_{\al\lambda} = \sum_{\al,\beta=1}^n \Big[\big(\CC\otimes I + I \otimes \CC\big)^m\Big]_{\mu\al\nu\beta}\, C_{\al\beta\lambda}
\end{equation}
for all $m\geq 1$. If $f(t)=\sum_{k=0}^\infty a_k t^k$ is an analytic function, then Eq. \eqref{2-9} yields
\begin{equation}\label{2-13}
\sum_{\al=1}^n C_{\mu\nu\al} f(\CC)_{\al\lambda} =\sum_{\al,\beta=1}^n f(\CC\otimes I + I\otimes \CC)_{\mu\al\nu\beta}\, C_{\al\beta\lambda}.
\end{equation}
Furthermore, by induction on $m$ one can show that
\begin{equation}
\frac{\p}{\p \p_\nu} (\CC^m)_{\mu\lambda} =\sum_{k=0}^{m-1} \sum_{\al,\beta=1}^n (\CC^k)_{\mu\al}\, C_{\al\nu\beta}\, (\CC^{m-1-k})_{\beta\lambda}, \quad m\geq 1.
\end{equation}
Using Eq. \eqref{2-9} we obtain
\begin{align}
\frac{\p}{\p \p_\nu} (\CC^m)_{\mu\lambda} &=\sum_{\al,\beta=1}^n \left[\sum_{k=0}^{m-1} (\CC^k\otimes I) (\CC\otimes I + I\otimes \CC)^{m-1-k}\right]_{\mu\al\nu\beta} C_{\al\beta\lambda} \notag \\
&=\sum_{\al, \beta=1}^n\left[\frac{(\CC\otimes I+I\otimes \CC)^m - \CC^m\otimes I}{I\otimes \CC}\right]_{\mu\al\nu\beta} C_{\al\beta\lambda}, \quad m\geq 1.  \label{2-11}
\end{align}
Relation \eqref{2-11} implies that for any analytic function $f(t)$ we have
\begin{equation}\label{2-14}
\frac{\p}{\p \p_\nu} f(\CC)_{\mu\lambda} =\sum_{\al,\beta=1}^n \left[\frac{f(\CC\otimes I+I\otimes \CC)-f(\CC\otimes I)}{I\otimes \CC}\right]_{\mu\al\nu\beta} C_{\al\beta\lambda},
\end{equation}
as well as
\begin{equation}\label{2-15}
\frac{\p}{\p \p_\nu} f(\CC)_{\mu\lambda} = -\sum_{\al,\beta=1}^n \left[\frac{f(\CC\otimes I + I\otimes \CC)-f(I\otimes \CC)}{\CC\otimes I}\right]_{\mu\al\nu\beta} C_{\al\beta\lambda}.
\end{equation}
We are now ready to prove our main result. We want to find $f(t)$ such that Eq. \eqref{2.4A} is a realization of the Lie algebra \eqref{2-1}.
The commutation relations for $\hat X_\mu$ imply that the functions $f(\CC)_{\mu\al}$ satisfy the following system of partial differential equations
\begin{equation}\label{2-17}
\sum_{\al=1}^n \left(f(\CC)_{\nu\al}\, \frac{\p}{\p \p_\al}f(\CC)_{\mu\lambda}-f(\CC)_{\mu\al}\, \frac{\p}{\p \p_\al} f(\CC)_{\nu\lambda}\right) =
\sum_{\al=1}^n C_{\mu\nu\al} f(\CC)_{\al\lambda}, \quad 1\leq \mu,\nu,\lambda \leq n.
\end{equation}
Using Eqs. \eqref{2-14} and \eqref{2-15} we find that
\begin{equation}\label{2-18}
\sum_{\al=1}^n f(\CC)_{\nu\al}\, \frac{\p}{\p \p_\al} f(\CC)_{\mu\lambda} = \sum_{\al,\beta=1}^n \left[f(I\otimes \CC)\frac{f(\CC\otimes I+I\otimes \CC)-f(\CC\otimes I)}{I\otimes \CC}
\right]_{\mu\al\nu\beta} C_{\al\beta\lambda}
\end{equation}
and also
\begin{equation}\label{2-19}
-\sum_{\al=1}^n f(\CC)_{\mu\al}\, \frac{\p}{\p \p_\al} f(\CC)_{\nu\lambda} = \sum_{\al,\beta=1}^n \left[f(\CC\otimes I)\frac{f(\CC\otimes I+I\otimes \CC)-f(I\otimes \CC)}{\CC\otimes I}
\right]_{\mu\al\nu\beta} C_{\al\beta\lambda}.
\end{equation}
Substituting Eqs. \eqref{2-13}, \eqref{2-18} and \eqref{2-19} into Eq. \eqref{2-17} it follows that $f(t)$  satisfies the functional equation
\begin{equation}\label{2-20}
f(t)\frac{f(t+s)-f(s)}{t}+f(s)\frac{f(t+s)-f(t)}{s} = f(t+s)
\end{equation}
where we have denoted $t=\CC\otimes I$ and $s=I\otimes \CC$. If we define $g(t)=1-t/f(t)$, then the functional equation \eqref{2-20} becomes $g(t)g(s)=g(t+s)$.
It is well known that the only solution is given by $g(t)=e^{a t}$, $a\in\R$, which implies
\begin{equation}
f(t) =\frac{t}{1-e^{at}}.
\end{equation}
The condition $f(0)=1$ holds only if $a=-1$, hence $f(t)$ is the generating function
for the Bernoulli numbers \eqref{Bernoulli}. Therefore, the obtained realization is given by
\begin{equation}\label{2.20A}
\hat X_\mu = \sum_{\al=1}^n x_\al \left(\frac{\CC}{I-e^{-\CC}}\right)_{\mu\al}.
\end{equation}
The realization \eqref{2.20A} was found in Ref. \cite{Durov} where several different proofs were presented using direct computation, formal geometry and
a coalgebra structure. In Ref. \cite{SKJ} it was shown that the realization \eqref{2.20A} can also be found by extending the original Lie algebra $\g_0$
by $n^2$ generators $T_{\mu\nu}$ satisfying the Lie bracket $[T_{\mu\nu},T_{\al\beta}]=0$ and $[T_{\mu\nu},X_\lambda]=\sum_{\al=1}^n C_{\mu\lambda\al} T_{\al\nu}$
in order to derive the function $f(t)$. Using an action of the Weyl algebra $\A_n$ on the space of polynomials $\K[x_1,\ldots, x_n]$, one can associate
an ordering prescription on $U(\g_0)$ to a given realization \eqref{2.3A} (see Ref. \cite{SKJ}). The realization \eqref{2.20A} corresponds to the Weyl
symmetric ordering on $U(\g_0)$, hence we refer to Eq. \eqref{2.20A} as the Weyl realization of $\g_0$.

We note that the functional equation \eqref{2-20} can also be solved by transforming it into a differential equation for
$f(t)$ or by using a recursion relation for the Taylor coefficients of $f(t)$. These alternative methods of solution are briefly outlined in the Appendix.

We remark that from equations \eqref{2-14} and \eqref{2.20A} the following commutators follow:
\begin{align}
[(e^{\CC})_{\mu\nu}, \hat X_\lambda] &=\sum_{\alpha=1}^n C_{\mu\lambda\alpha} (e^{\CC})_{\alpha\nu}, \\
[(e^{-\CC})_{\mu\nu}, \hat X_\lambda] &= -\sum_{\alpha=1}^n C_{\alpha\lambda\nu} (e^{-\CC})_{\mu\alpha}.
\end{align}
The operators $(e^{\CC})_{\mu\nu}$ and $(e^{-\CC})_{\mu\nu}$ correspond to the left and right shift operators $T_{\mu\nu}$ and $S_{\mu\nu}$ respectively introduced in Ref. \cite{SKJ}.

\section{Generalization to a certain type of Lie superalgebras}
\label{sec-3}

The proof of the Weyl symmetric realization \eqref{2.20A} presented in Section \ref{sec-2} can be extended to a certain type of Lie superalgebras
defined as follows. Let $\g_0$ and $\g_1$ be vector spaces over $\K$ with bases $X_1, X_2, \ldots, X_n$ and $\theta_1, \theta_2, \ldots, \theta_m$, respectively.
Define the $\Z_2$--graded vector space $\g=\g_0\oplus \g_1$ with degrees of homogeneous elements $\bar X=0$ if $X\in \g_0$ and $\bar X=1$ if $X\in \g_1$.
In the rest of the paper it is convenient to adopt the following notation. The lowercase greek letters $\alpha, \beta, \gamma$, etc. range over the indices of $X_1, X_2, \ldots, X_n$
and lowercase latin letters $a,b,c$, etc. range over the indices of $\theta_1, \theta_2, \ldots, \theta_m$. Define the Lie superbracket on $\g$ by
\begin{equation}\label{3.23}
[X_\mu,X_\nu]=\sum_{\al=1}^n C_{\mu\nu\al} X_\al, \quad [\theta_a, X_\nu]=\sum_{b=1}^m K_{a\nu b} \theta_b, \quad [\theta_a, \theta_b]=0.
\end{equation}
Recall that the Lie superbracket is graded skew--symmetric, $[X,Y]=-(-1)^{\bar X \bar Y} [Y,X]$, and it satisfies the graded Jacobi identity
\begin{equation}\label{g-Jacobi}
(-1)^{\bar X \bar Z} [X,[Y,Z]]+(-1)^{\bar Y \bar X} [Y,[Z,X]]+(-1)^{\bar Z \bar Y} [Z,[X,Y]]=0
\end{equation}
for all $X,Y,Z\in \g$. This condition implies that the structure constants $C_{\mu\nu\al}$ and $K_{a\nu b}$ satisfy
\begin{align}
&\sum_{\rho=1}^n \big(C_{\mu\al\rho} C_{\rho\beta\nu}+C_{\al\beta\rho} C_{\rho\mu\nu}+C_{\beta\mu\rho} C_{\rho\al\nu}\big) = 0, \label{3.25}  \\
&\sum_{b=1}^m \big(K_{a \nu b} K_{b\mu c}-K_{a \mu b} K_{b \nu c}\big) + \sum_{\rho=1}^n C_{\mu\nu\rho} K_{a \rho c} = 0.  \label{3.26}
\end{align}
In the enveloping algebra $U(\g)$ the generators $X_\mu$ and $\theta_a$ are subject to relations
\begin{align}
X_\mu X_\nu - X_\nu X_\mu &= \sum_{\al=1}^n C_{\mu\nu\al} X_\al,  \label{3.27}\\
\theta_a X_\mu - X_\mu \theta_a &= \sum_{b=1}^m K_{a\mu b} \theta_b,  \label{3.28}\\
\theta_a \theta_b + \theta_b \theta_a &= 0,  \label{3.29}
\end{align}
where for simplicity we identify $X_\mu$ and $\theta_a$ with their canonical images in $U(\g)$. The enveloping algebra $U(\g)$ has applications
in the construction of bicovariant differential calculus on NC spaces of the Lie algebra type \cite{SKJ-FODC}. As remarked earlier, $X_1, X_2, \ldots, X_n$ are
interpreted as NC coordinates satisfying the Lie algebra type commutation relations \eqref{2-1}, and $\theta_1, \theta_2, \ldots, \theta_m$ are interpreted as
anticommuting one--forms on the NC space $U(\g_0)$. We note that in the noncommutative setting the number of one--forms is not necessarily equal to the number
of coordinates. An example of such calculus on $\kappa$--Minkowski space can be found in Ref. \cite{Sitarz}. For more details on the mathematical
properties and physical applications of differential calculus on NC spaces see Refs.
\cite{Woronowicz, Vladimirov, Schupp, Landi, Sitarz, Meljanac-1, SKJ-FODC, KS, Meljanac-2, Meljanac-3, Meljanac-4}.

The commutation relations \eqref{3.27}--\eqref{3.29} can be written in a more compact form if we introduce the following convention. Let the uppercase latin letters $A,B,C$, etc. range
over both the lowercase greek and latin letters. With this convention $\sum_A = \sum_{\al=1}^n + \sum_{a=1}^m$. Denote the generators of $U(\g)$ by $Z_1, Z_2, \ldots,
Z_{n+m}$ where $Z_\al = X_\al$ and $Z_{n+a}=\theta_a$. Then the commutation relations \eqref{3.27}--\eqref{3.29} can be written as
\begin{equation}\label{2.29A}
[Z_A,Z_B]=\sum_J \widetilde C_{ABJ} Z_J
\end{equation}
where $[Z_A,Z_B]=Z_A Z_B - (-1)^{\bar Z_A \bar Z_B} Z_B Z_A$ and the structure constants $\widetilde{C}_{ABJ}$ are given explicitly by
\begin{alignat}{2}
&\widetilde C_{\mu\nu\lambda} = C_{\mu\nu\lambda}, \qquad &  &\widetilde C_{\mu\nu a} =0, \label{C-18} \\
&\widetilde C_{\mu a\nu} =0, \qquad &  &\widetilde C_{\mu ab} =-K_{a\mu b}, \\
&\widetilde C_{a \mu\nu} =0, \qquad & &\widetilde C_{a\mu b} = K_{a\mu b}, \\
&\widetilde C_{a b\mu} = 0, \qquad & &\widetilde C_{abc} = 0. \label{C-21}
\end{alignat}
For this type of Lie superalgebras, relations \eqref{3.25}--\eqref{3.26} are equivalent with the single Jacobi identity without grading
\begin{equation}\label{3.33}
\sum_J (\widetilde{C}_{ABJ}\, \widetilde{C}_{JCD}+\widetilde{C}_{BCJ}\,\widetilde{C}_{JAD}+\widetilde{C}_{CAJ}\, \widetilde{C}_{JBD})=0
\end{equation}
and the structure constants $\widetilde{C}_{ABJ}$ are skew--symmetric in the first two indices, $\widetilde{C}_{ABJ}=-\widetilde{C}_{BAJ}$.

We now consider a realization of the Lie superalgebra \eqref{3.23} by formal power series in a semicompleted Weyl superalgebra (Clifford--Weyl algebra)
analogous to that given by Eq. \eqref{2.20A}. Let $\A_{(n,m)}$ denote the Weyl superalgebra generated by four types of generators $x_\mu, \p_\mu, \xi_a, q_a$,
$1\leq \mu\leq n$, $1\leq a \leq m$, with degrees defined by $\bar x_\mu = \bar \p_\mu=0$ and $\bar \xi_a = \bar q_a=1$. The algebra $\A_{(n,m)}$ is equipped with the
supercommutator $[x,y]=xy-(-1)^{\bar x \bar y} y x$ and the generators are subject to relations
\begin{equation}
[\p_\mu,x_\nu]=\delta_{\mu\nu}, \quad [q_a,\xi_b]=\delta_{ab}
\end{equation}
with all other relations between the generators being zero. Let us denote the generators of $\A_{(n,m)}$ by $D_A$ and $z_A$ where $D_\al = \p_\al$, $D_{n+a}=q_a$,
$z_\al =x_\al$ and $z_{n+a}=\xi_a$. Define the $(n+m)\times (n+m)$ matrix $\wtbfC_{AB} = \sum_J \widetilde C_{AJB} D_J$. Since the structure constants
$\widetilde C_{ABJ}$ satisfy the same formal properties as the structure constants $C_{\mu\nu\al}$ of the Lie algebra $\g_0$ (and the generators $D_A$ pairwise commute),
by repeating the argument from Section \ref{sec-2} one can show that
\begin{equation}\label{3.35}
\sum_J \widetilde C_{ABJ} (\widetilde{\mathbf{C}}^k)_{JC} = \sum_{J,K} \big[(\widetilde{\mathbf{C}}\otimes I + I \otimes \widetilde{\mathbf{C}})^k\big]_{AJBK}\, \widetilde C_{JKC},
\quad k\geq 1,
\end{equation}
where $I$ is the $(n+m)\times (n+m)$ identity matrix. Thus, for an analytic function $f(t)=\sum_{k=0}^\infty a_k t^k$ we have
\begin{equation}\label{3.37}
\sum_J \widetilde C_{ABJ} f(\widetilde{\mathbf{C}})_{JC}=\sum_{J,K} f(\widetilde{\mathbf{C}}\otimes I + I \otimes \widetilde{\mathbf{C}})_{AJBK}\, \widetilde C_{JKC}.
\end{equation}
Let us define the graded derivative
\begin{equation}
\tilde \p_A h(z)\equiv [h(z),z_A]
\end{equation}
where $[\cdot , \cdot ]$ is the supercommutator in $\A_{(n,m)}$. Then from the Leibniz rule it follows that
\begin{equation}
\tilde \p_B (\widetilde{\mathbf{C}}^n)_{AC} = \sum_{k=0}^{n-1} (\widetilde{\mathbf{C}}^k)_{AJ}\, \widetilde C_{JBK}\, (\widetilde{\mathbf{C}}^{n-1-k})_{KC}.
\end{equation}
Using Eq. \eqref{3.35} this becomes
\begin{align}
\tilde \p_B (\widetilde{\mathbf{C}}^n)_{AC} &= \sum_{J,K} \Big[\sum_{k=0}^\infty (\widetilde{\mathbf{C}}^k\otimes I)(\widetilde{\mathbf{C}}\otimes I + I \otimes
\mathbf{\widetilde{C}})^{n-1-k}\Big]_{AJBK} \widetilde C_{JKC} \notag \\
&= \sum_{J,K} \Big[\frac{(\widetilde{\mathbf{C}}\otimes I + I \otimes \widetilde{\mathbf{C}})^n-\widetilde{\mathbf{C}}^n \otimes I}{I\otimes \widetilde{\mathbf{C}}}
\Big]_{AJBK} \widetilde C_{JKC}  \label{3.39}
\end{align}
which readily generalizes to
\begin{equation}\label{3.41}
\tilde\p_B f(\widetilde{\mathbf{C}})_{AC} =  \sum_{J,K} \Big[\frac{f(\widetilde{\mathbf{C}}\otimes I + I \otimes \widetilde{\mathbf{C}})-f(\widetilde{\mathbf{C}} \otimes I)}{I\otimes \widetilde{\mathbf{C}}}
\Big]_{AJBK} \widetilde C_{JKC}.
\end{equation}
We want to find an analytic function $f(t)$ satisfying $f(0)=1$ such that
\begin{equation}\label{3.41A}
\hat Z_A = \sum_J z_J f(\wtbfC)_{AJ}
\end{equation}
is a realization of the Lie superalgebra \eqref{2.29A}. By substituting Eq. \eqref{3.41A} into Eq. \eqref{2.29A}
we find
\begin{equation}\label{3.42}
\sum_J f(\wtbfC)_{BJ}\, \tilde\p_J f(\wtbfC)_{AC} - f(\wtbfC)_{AJ}\, \tilde \p_J f(\wtbfC)_{BC} = \sum_J \widetilde C_{ABJ} f(\wtbfC)_{JC}.
\end{equation}
Furthermore, from Eq. \eqref{3.41} it follows that
\begin{equation}\label{3.43}
\sum_J f(\wtbfC)_{BJ}\, \tilde \p_J f(\wtbfC)_{AC} = \sum_{J,K} \left[f(I\otimes \wtbfC) \frac{f(\wtbfC\otimes I + I \otimes \wtbfC)-f(\wtbfC\otimes I)}{I\otimes \wtbfC}\right]_{AJBK}
\widetilde C_{JKC}
\end{equation}
and also
\begin{equation}\label{3.44}
\sum_J f(\wtbfC)_{AJ} \, \tilde\p_J f(\wtbfC)_{BC} = - \sum_{J,K} \left[f(\wtbfC\otimes I) \frac{f(\wtbfC\otimes I+I\otimes \wtbfC)-f(I\otimes \wtbfC)}{\wtbfC\otimes I}\right]_{AJBK}
\widetilde C_{JKC}.
\end{equation}
Substituting Eqs. \eqref{3.43}--\eqref{3.44} and Eq. \eqref{3.37} into Eq. \eqref{3.42} shows that $f(t)$ satisfies the functional equation \eqref{2-20} with
$t=\wtbfC\otimes I$ and $s=I\otimes \wtbfC$. As proved in Section \ref{sec-2}, this equation has a unique solution $f(t)=t/(1-e^{-t})$ such that $f(0)=1$. Hence,
\begin{equation}\label{3.45}
\hat Z_A = \sum_J z_J \left(\frac{\wtbfC}{1-e^{-\wtbfC}}\right)_{AJ}
\end{equation}
is a realization of the Lie superalgebra \eqref{2.29A}, i.e. the Lie superalgebra defined by Eq. \eqref{3.23}.
Relation \eqref{3.45} is completely analogous to the Weyl realization \eqref{2.20A} of the Lie algebra $\g_0$. It represents a generalization of this
realization to Lie superalgebras given by Eq. \eqref{3.23}.

Now let us examine explicitly expression \eqref{3.45}.
Due to the specific form of the structure constants $\widetilde C_{ABJ}$, the matrix $\wtbfC$ is upper triangular,
\begin{equation}
\widetilde{\mathbf{C}}=\begin{bmatrix}\CC & \mathbf{K} \\ \mathbf{0} & \mathbf{L}\end{bmatrix}
\end{equation}
where $\CC$ is defined as before, $\mathbf{L}$ is the $n\times m$ matrix $\mathbf{L}_{\mu a} = -\sum_{b=1}^m K_{b\mu a} q_b$ and $\KK$ is the
$m\times m$ matrix given by $\KK_{ab} = \sum_{\rho=1}^n K_{a\rho b} \p_\rho$. If $f(t)$ is the generating function for the Bernoulli numbers \eqref{Bernoulli},
then the matrix $f(\widetilde{\mathbf{C}})$ is given by
\begin{equation}\label{3.47}
f(\widetilde{\mathbf{C}})=\begin{bmatrix} f(\CC) & \mathbf{F} \\ \mathbf{0} & f(\KK) \end{bmatrix}
\end{equation}
where
\begin{equation}\label{3.48}
\mathbf{F} = \sum_{k=1}^\infty \sum_{l=1}^k \frac{(-1)^k}{k!} B_k \CC^{k-l}\, \mathbf{L}\, \KK^{l-1}.
\end{equation}
Hence, using equations \eqref{3.47} and \eqref{3.48} we find that the realizations of the generators $X_\mu$ and $\theta_a$ are explicitly given by
\begin{align}
\hat X_\mu &= \sum_{\al=1}^n x_\al f(\wtbfC)_{\mu\al}+\sum_{a=1}^m \xi_a f(\wtbfC)_{\mu a} \notag \\
&=\sum_{\al=1}^n x_\al \left(\frac{\CC}{I-e^{-\CC}}\right)_{\mu\al} + \sum_{k=1}^\infty \sum_{l=1}^k \sum_{a=1}^m \frac{(-1)^k}{k!} B_k\, \xi_a\,
(\CC^{k-l}\, \mathbf{L}\, \KK^{l-1})_{\mu a}
\end{align}
and
\begin{equation}
\hat \theta_a = \sum_{\beta=1}^n x_\beta f(\wtbfC)_{\beta a} + \sum_{b=1}^m \xi_b f(\wtbfC)_{ab} =\sum_{b=1}^m \xi_b \left(\frac{\KK}{I-e^{-\KK}}\right)_{ab}.
\end{equation}
This proves the conjecture stated in Ref. \cite{SKJ-FODC}. We point out that the proof presented here applies only to Lie superalgebras $\g=\g_0\oplus \g_1$ with $[\g_1,\g_1]=\{0\}$.
In future work we shall investigate if this approach can be extended to all types of Lie superalgebras.

Note that from equations \eqref{3.41} and \eqref{3.45} the following commutators follow:
\begin{align}
[(e^{\wtbfC})_{AB}, \hat Z_C] &= \sum_{D}\widetilde C_{ACD}(e^{\wtbfC})_{DB}, \\
[(e^{-\wtbfC})_{AB}, \hat Z_C] &= -\sum_{D} \widetilde  C_{DCB} (e^{-\wtbfC})_{AD}.
\end{align}
The operators $(e^{\wtbfC})_{AB}$ and $(e^{-\wtbfC})_{AB}$ correspond to the left and right shift operators $T_{AB}$ and $S_{AB}$ respectively introduced in Ref. \cite{SKJ-FODC}.

\section*{Acknowledgements}
The work by S.M. and D.P. has been supported by Croatian Science Foundation under the Project No. IP-2014-09-9582 as well as by the H2020 Twinning project No. 692194, ``RBI-T-WINNING''.

\section*{Appendix: Alternative solutions of the functional equation}

In this Appendix we sketch alternative methods for obtaining a solution of the functional equation \eqref{2-20}.\\

\noindent\textbf{A1. Differential equation}

Recall that the classical limit condition $\lim_{h\to 0} \hat x_\mu = x_\mu$ holds if $f(0)=1$. Taking the limit as $s\to 0$ in both sides of Eq. \eqref{2-20}
we obtain the initial value problem
\begin{equation}\label{A-23}
f(t)\frac{f(t)-1}{t}+f^\prime (t) = f(t), \quad f(0)=1.
\end{equation}
Introducing the function $g(t)=f(t)/t$ we find that
\begin{equation}\label{A-24}
g^\prime (t) = g(t) (1-g(t)).
\end{equation}
The solution of equation \eqref{A-24} is given by
\begin{equation}
g(t) = \frac{1}{1+K e^{-t}}
\end{equation}
where $K$ is a constant of integration. Thus,
\begin{equation}
f(t)=\frac{t}{1+K e^{-t}}.
\end{equation}
The condition $\lim_{x\to 0} f(t)=1$ is satisfied only if $K=-1$ since $\lim_{x\to 0} f(t)=0$ for $K\neq -1$. Therefore, the solution of Eq. \eqref{A-23}
is given by $f(t)=t/(1-e^{-t})$. \\

\noindent\textbf{A.2 Recursion relation}

We look for the solution of the initial value problem \eqref{A-23} in the form of an analytic function
$f(t)=\sum_{k=0}^\infty b_k t^k$. The condition $f(0)=1$ implies $b_0=1$. Substituting the series expansion for $f(t)$ into Eq. \eqref{A-23} we obtain
a recursion relation for $b_n$,
\begin{equation}\label{A-27}
\sum_{k=0}^{n-1} b_k b_{n-k} + n b_n = b_{n-1}, \quad b_0=1, \quad n\geq 1.
\end{equation}
Solving Eq. \eqref{A-27} for $n=1,2,3$ we find $b_1=1/2$, $b_2=1/12$ and $b_3=0$. We claim that $b_n=0$ for all odd $n\geq 3$. This is easily shown by
induction on $n$. Suppose that $b_k=0$ for all odd $k<n$ where $n\in \N$ is some odd number. Note that
\begin{equation}\label{A-28}
\sum_{k=0}^{n-1} b_k b_{n-k} = b_0 b_n + \sum_{k=2}^{n-2} b_k b_{n-k} + b_{n-1} b_1 = b_n + b_{n-1} + \sum_{k=2}^{n-2} b_k b_{n-k}
\end{equation}
because $b_0=1$ and $b_1=1/2$. Substituting Eq. \eqref{A-28} into Eq. \eqref{A-27} we find
\begin{equation}\label{A-29}
(n+1) b_n = -\sum_{k=2}^{n-2} b_k b_{n-k}.
\end{equation}
Since $n$ is odd, either $k$ or ${n-k}$ is odd in the sum in Eq. \eqref{A-29}. Thus, by the induction hypothesis $b_k=0$ or $b_{n-k}=0$ which implies
$\sum_{k=2}^{n-2} b_k b_{n-k} = 0$. Therefore, $b_n=0$ proving that $b_n=0$ for all odd $n\geq 3$.

In the next step we show that $b_n$ are related to the Bernoulli numbers $B_n$. It suffices to consider the even indices $n=2m$. In this case relation
\eqref{A-27} becomes
\begin{equation}\label{A-30}
\sum_{k=2}^{2m-1} b_k b_{2m-k} + 2m b_{2m} = 0, \quad m\geq 2,
\end{equation}
since $b_{2m-1}=0$. Taking into account that $b_{2m+1}=0$ for all $m\geq 1$, the sum in Eq. \eqref{A-30} contains only the terms with even indices, hence
\begin{equation}
\sum_{k=0}^{2m-1} b_k b_{2m-k} = \sum_{k=0}^{m-1} b_{2k} b_{2m-2k}.
\end{equation}
Therefore, it follows from Eq. \eqref{A-30} that
\begin{equation}\label{A-32}
\sum_{k=0}^{m-1} b_{2k} b_{2m-2k} + 2m b_{2m} =0, \quad m\geq 2.
\end{equation}
It is well known that the Bernoulli numbers $B_{2k}$ satisfy the recursion relation
\begin{equation}
\sum_{k=0}^{m-1}\binom{2m}{2k} B_{2k} B_{2m-2k} + 2m B_{2m} =0
\end{equation}
which is equivalent with
\begin{equation}\label{A-34}
\sum_{k=0}^{m-1} \frac{B_{2k}}{(2k)!} \frac{B_{2m-2k}}{(2m-2k)!} + 2m \frac{B_{2m}}{(2m)!}=0.
\end{equation}
Comparing Eqs. \eqref{A-34} and \eqref{A-32} we conclude that $b_k = B_k/k!$ for $k\geq 0$ (with convention $B_1=1/2$) where we have also taken into
account that $b_{2k+1}=B_{2k+1}=0$ for all $k\geq 1$. Therefore, the solution of Eq. \eqref{A-23} is given by
\begin{equation}\label{A-35}
f(t) =\sum_{k=0}^\infty \frac{B_k}{k!} t^k = \frac{t}{1-e^{-t}}.
\end{equation}
We remark that the solution \eqref{A-35} is sometimes in the literature also written as\\ $\sum_{k=0}^\infty (-1)^k (B_k/k!) t^k$ with convention $B_1=-1/2$.

\end{document}